\documentclass[10pt,journal,compsoc]{IEEEtran}
\usepackage{tipa}
\usepackage{algorithm}
\usepackage[noend]{algpseudocode}
\usepackage{pifont}
\usepackage{algorithmicx}
\usepackage{amsmath,amssymb,amsthm,mathrsfs,amsfonts,dsfont}
\usepackage{url}
\usepackage{xcolor}
\usepackage{pgfplots}
\usepackage{tcolorbox}
\usepackage{tikz}
\usepackage{hyperref}
\hypersetup{
    colorlinks=true,
    linkcolor=blue,
    filecolor=magenta,      
    urlcolor=cyan,
}

\algnewcommand\algorithmicinput{\textbf{Input:}}
\algnewcommand\Input{\item[\algorithmicinput]}
\algnewcommand\algorithmicoutput{\textbf{Output:}}
\algnewcommand\Output{\item[\algorithmicoutput]}

\ifCLASSOPTIONcompsoc
  \usepackage[nocompress]{cite}
\else
  \usepackage{cite}
\fi
\ifCLASSINFOpdf
\else
\fi
\begin{document}
\bstctlcite{IEEEexample:BSTcontrol}
\title{So you want to be a Super Researcher?}

\author{Sanjay Rathee$^*$ and Sheah Lin Lee$^*$ % <-this % stops a space
\IEEEcompsocitemizethanks{\IEEEcompsocthanksitem S. Rathee was with the Department of Oncology, University of Oxford, Oxford, United Kingdom, OX37DQ.\protect\\
E-mail: sanjaysinghrathi@gmail.com
\IEEEcompsocthanksitem S.L. Lee was with Cancer Sciences Unit, Faculty of Medicine, University of Southampton.\protect\\
E-mail: s.l.lee@soton.ac.uk}% <-this % stops an unwanted space
\thanks{Manuscript received }}


\markboth{Journal of \LaTeX\ Class Files,~Vol.~14, No.~8, August~2015}%
{Shell \MakeLowercase{\textit{et al.}}: Bare Demo of IEEEtran.cls for Computer Society Journals}



\IEEEtitleabstractindextext{%
\begin{abstract}
Publishing original scientific research is inherent to the work of a researcher. However, the pressure to maintain productivity and scientific impact can lead to research group publishing excessively, negatively affecting the mental health of a researcher. Ph.D. students and early career researchers are particularly susceptible to this pressure due to the inherent vulnerability of their positions. At present, there are no resources that concisely summarise the publication culture of a research group to help the researcher make an informed decision before joining. In this article, we present the 'Super Researcher' app, an R Shiny application(app) with a user-friendly interface. Using text-mining methodology to extract publicly available author data from Scopus, this pilot app has four fundamental functions to provide snapshot information that will help researchers grasp the publication culture of a research group within minutes. The 'Super Researcher' app provides information on: 1) institution data, 2) author's publication, 3) co-author network plots and 4) publication journals.

The 'Super Researcher' app is built on R shiny which provides an interactive interface to users. This app utilizes the Big Data framework Apache Spark to mine relevant information from a huge author information database. The author's information is stored and manipulated using both SQL(SQLite) and NoSQL(HBase) databases. Hbase is used for local data storage and manipulation while SQLite feeds data to the R Shiny interface. 

In this paper, we introduce these functionalities and illustrate how this information can help guide a researcher to select a new Principle Investigator (PI) with better compatibility in terms of publication attitude using a case study.\emph{}  	
Available: https://researchmind.co.uk/super-researcher/
\end{abstract}

% Note that keywords are not normally used for peer review papers.
}

\maketitle

\IEEEdisplaynontitleabstractindextext

% \fi
\IEEEpeerreviewmaketitle

\IEEEraisesectionheading{\section{Introduction}\label{sec:introduction}}

\IEEEPARstart{P}{ublication} culture is a broad term that takes into account the number of publications, the impact of the publications, and the general productivity of a research group and network. Publication is inherent to the work of a researcher. In fact, it is the duty of a researcher to publish original, insightful, and significant scientific findings, which positively contribute to the pursuit of knowledge. In reality, the competitive nature of academia often transformed the purpose of publication from 'knowledge dissemination' to a 'measurement of success'. The number of publications and the amount of grant monies accrued is often used to quantify productivity and scientific impact, comparing one scientist to another, and can affect employability. In addition, institutions are incentivised to use the number of publications as part of their assessment frameworks to evaluate faculty members' productivity ~\cite{coriat2019phd}. As the number of publications is often used as the 'hard' currency of academia, it has a knock-on effect on the publication culture of a research group.

Many researchers welcome the opportunity to publish. However, the pressure to publish may lead to ‘excessive publishing’. The pressure to publish and its effect on science and the scientific community has been increasingly discussed in recent years. On one hand, it exacerbates the mental health epidemic that is affecting the scientific communities, a plight which has been highlighted recently in PhD students and early career researchers survey conducted by Nature ~\cite{woolston2019phds,woolston2020postdocs}. On the other hand, and perhaps more detrimental, the pressure to publish can push down the quality of research in the long term, as discussed by Daniel Sarewitz ~\cite{sarewitz2016pressure}. Reproducible scientific work takes time, sometimes years, to produce. Yet, no one has openly acknowledge how many publications in a year is too many for scientific trustworthiness to be called into question, or indeed, its effect on the mental health of a researcher.

More often than not, publications are written up by postgraduates (PhD or MSc students) and early career researchers within a research group. For a new researcher, the publication culture may not be at the top of their priority when looking for a new research group and a principle investigator (PI) to be their mentor and guide for the next few years. Prospective researchers may relish at joining a position with ample of opportunities to publish. However, numerous anecdotal incidences have highlighted that balancing the pressure to publish excessive amount of research papers whilst conducting high quality research is an overwhelming experience and detrimental to mental health. This harsh reality, grimly embodied by the phrase "publish or perish" as coined by Coolidge in 1932 ~\cite{coolidge1932archibald}, is often realised too late by hopeful researchers who are already committed into a student programme or an employment contract. On the contrary, there remains a handful of researchers who would like to join a research group which publishes high volume of articles to boost their curricula vitae. 

Whatever the intent, in this paper, we present the 'Super Researcher' application (app) to help researchers make informed decisions about the publication culture of a research group. Using text-mining methods and publication data from Scopus, the 'Super Researcher' app enables easy browsing of an institution and individual research group's publication culture. The 'Super Researcher' app concisely summarises all publications of an author, either as a first, middle or last author, and the total and average citation per publication per year. Crucially, it also provides information on an author's network of co-authors, giving a broader view on the publication culture in his or her research group. Lastly, the app provides a breakdown of scientific journals that said author publishes in. To our knowledge, the 'Super Researcher' app is the first app which addresses the publication culture of an institute and individual research groups.

\section{Super Researcher - the application}
The 'Super Researcher' app is an R Shiny app with a user-friendly interface. This pilot app has four fundamental functions to present information that aids researchers in understanding the publication culture of the potential new research group and its host institution. 

As a researcher who is choosing a new research group to join, the assumption is that he or she would be seeking an opportunity to be the first author in publications. Authorship in the 'Super Researcher' app is classified into 'first author', 'middle author', and 'last author'. This classification reflects the work and effort needed to be the first author in any publication, a task that is usually carried out by postgraduate or early career researchers and takes up a significant proportion of their work time.

\subsection{The Cutoff}
Who is a 'Super Researcher'? In the 'Super Researcher' app, the primary aim is to address the publication culture. Hence, we have narrowly defined a 'Super Researcher' as a researcher who has published a set minimum amount of first author publications per year as a 'Super Researcher'. This set minimum amount is referred to as 'cutoff', which is an integral part of the app. 

The 'cutoff' is not a set but a dynamic value that can be altered. Although we suggest a cutoff of more than five first-author publications per year (an average of one first-author publication every two months) as the cutoff for 'Super Researcher', it is up to the discretion of the user to decide how many publications per year as the first author is an acceptable workload for them. The 'cutoff' is set at '0' by default.

\subsection{Institution Information}
Although the opportunity for authorship is an important factor, the geographical area, prestige and reputation attached to a said institute play an important role in the decision-making process when choosing a research group to join. Therefore, the 'Super Researcher' app first provides an overview of the publication culture in an institute. Once the minimum cutoff value is set, the app will extract and present all researchers in the said institute who are classified as 'Super Researcher'. This overview of an institute or a faculty can help identify like-minded research groups and guide further enquiry. 

\begin{figure}[!h]
\includegraphics[height=1.2in, width=3.5in]{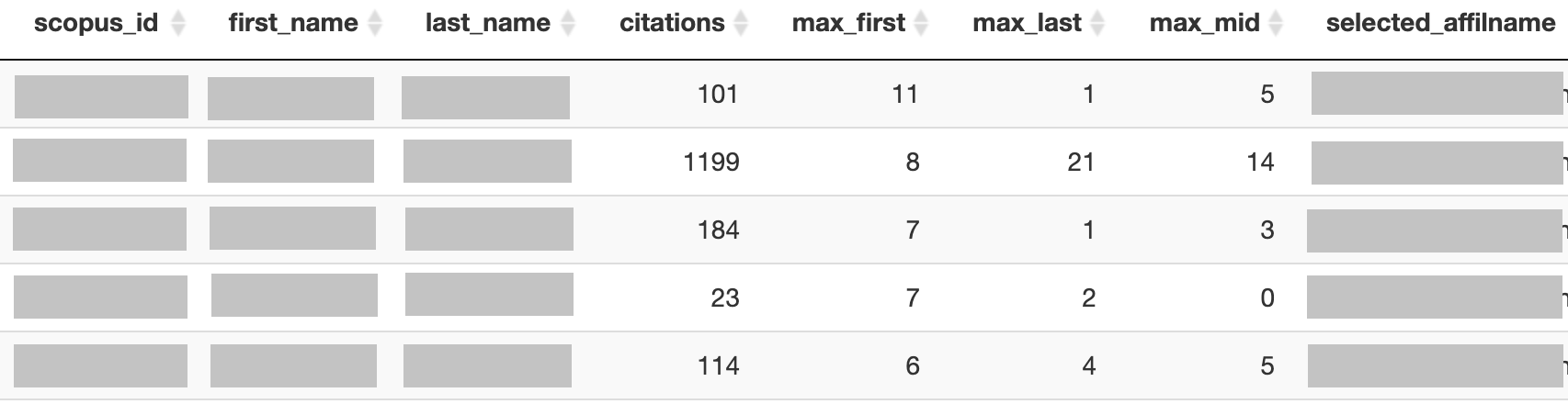}
\caption{Screenshot of table presenting individual author summary in a said institute from the 'Super Researcher' app.}
\label{fig:institute1}
\end{figure}

The app allows queries on the individual institution or multiple institutions in the same query. Once selected, the app generates a table and graph. The table presents information on all authors from the selected institutions who are present on the Scopus database. For each author, a Scopus ID is provided, together with their names, number of citations, the maximum number of first author (criteria of a 'Super Researcher'), last author and mid author publications in one year. For example, in Figure~\ref{fig:institute1}, the first entry shows an author in the selected institute who had published a maximum number of 11 first author publications, 1 last author publication and 5 mid author publications in any year. It also provides the highest number of citations in any year. For a further breakdown of the trend of these publication figures, we recommend using the 'individual author' feature of the app. 

\begin{figure}[!h]
\includegraphics[height=2.3in, width=3.5in]{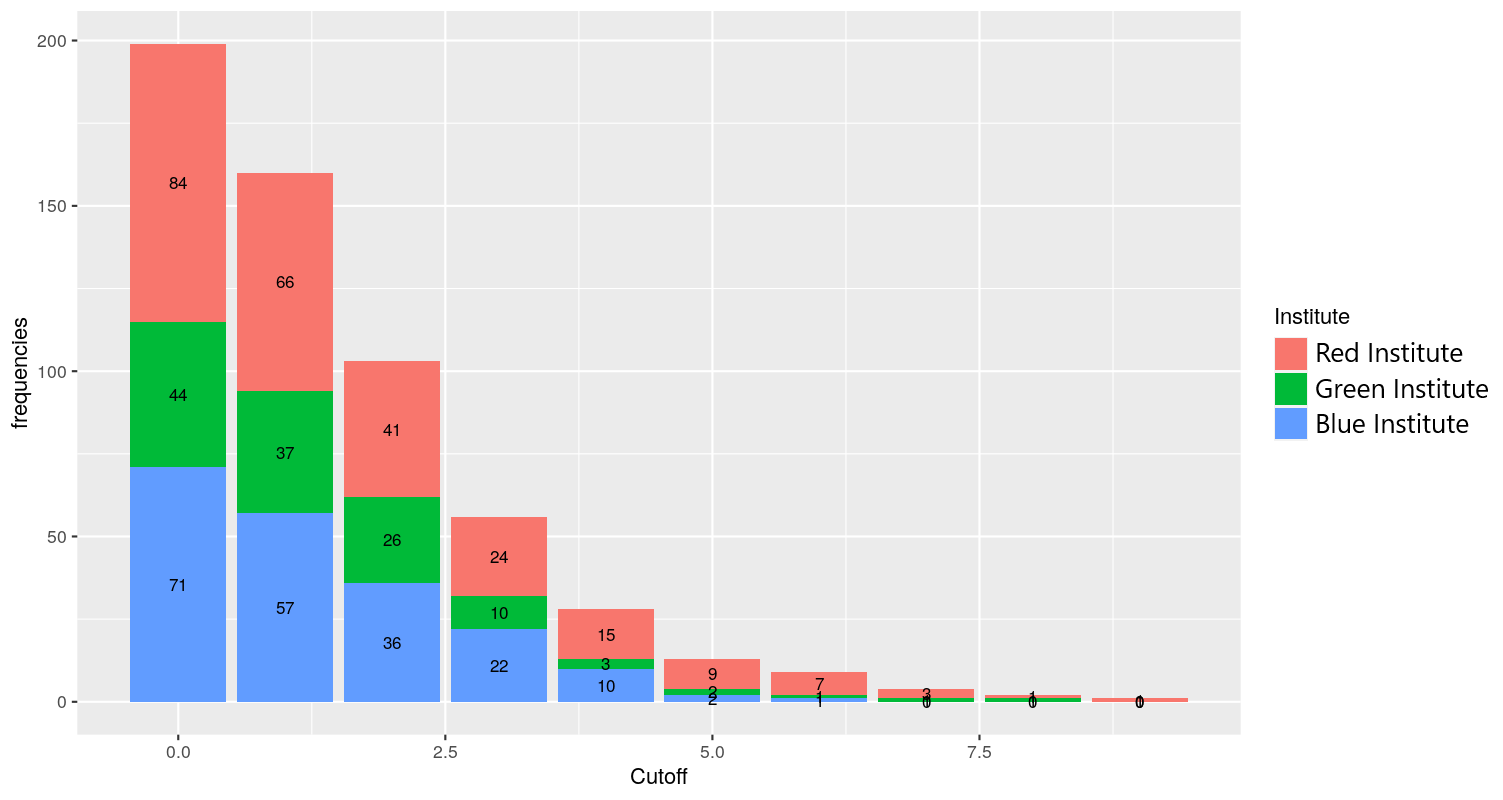}
\caption{Barplot showcasing the publication culture of three institutes.}
\label{fig:institute2}
\end{figure}

Figure~\ref{fig:institute2} shows the barplot which summarises the presence of 'Super Researchers' in three institutes. The horizontal axis indicates the maximum number of first-author publications per year for each author, and the vertical axis indicates the number of authors. In figure 2, there are 84 authors who had published at least one first-author article (cutoff more than 0) in any year in Red institute. Out of these 84 authors, 9 (11 \%) published at least 6 first-author publications in any year. The institution barplot gives an overview of the publication culture in the institute. We expect most institutes to have a hyperbolic curve with a short tail. The presence of a hyperbolic curve with a long tail indicated the presence of a small number of researchers who are publishing excessively. More alarming is the presence of a flattened hyperbola with a long tail, which suggests that a high proportion of researchers in the institute is publishing excessively.     

\subsection{Individual author}
\begin{figure}[!h]
\includegraphics[height=2.3in, width=3.5in]{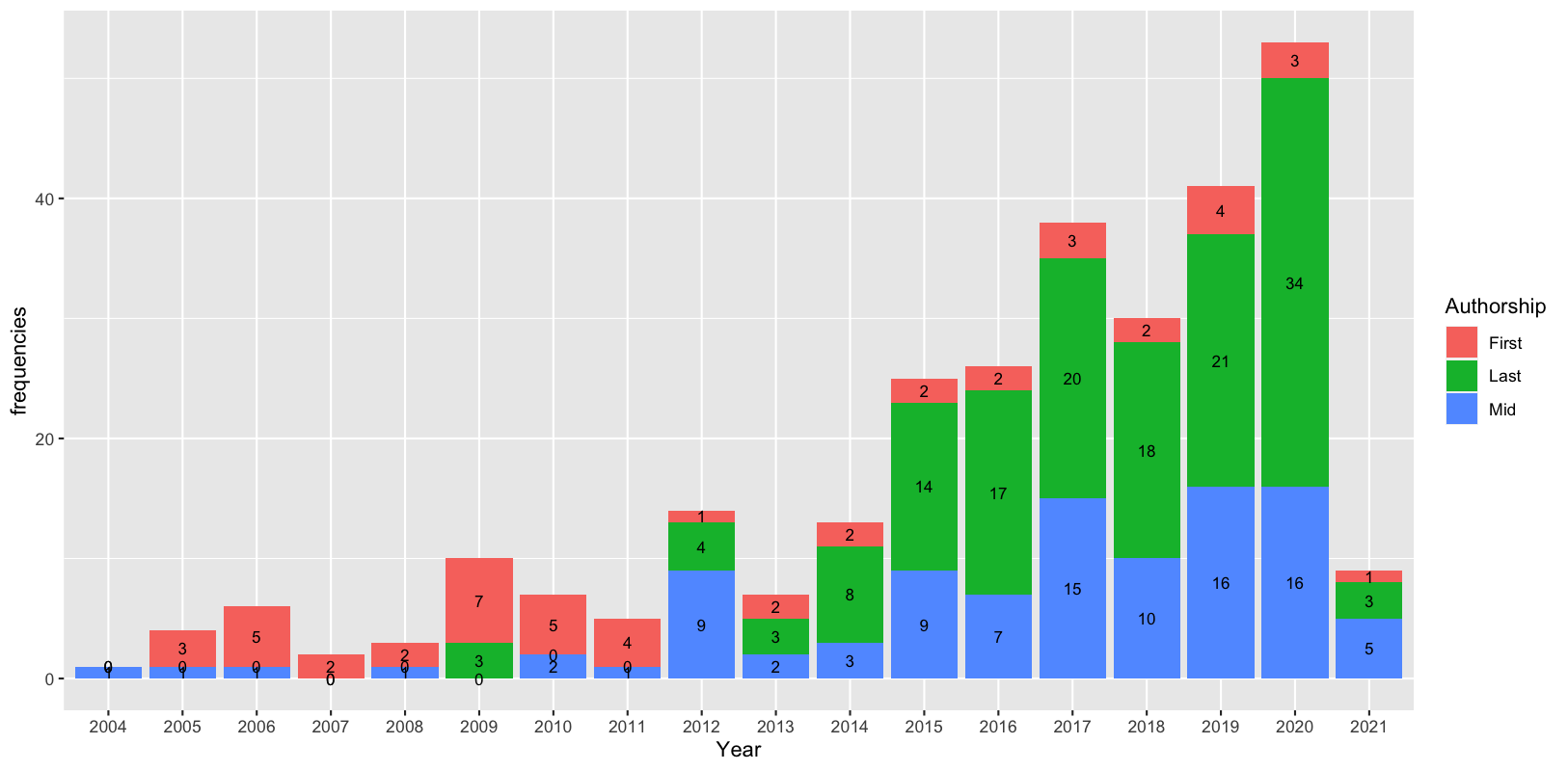}
\caption{Number of publications per year by a 'Super Researcher' stratified by order of authorship}
\label{fig:read1}
\end{figure}

The next section focuses on information about an individual author. Users can search for an individual author using their first and last name or their unique Scopus ID. The user can enter a 'cutoff' value, or leave it as '0' to get all information about the author. The app will present a table which includes the number of citations, number of first, mid and last author publications for each year that the author fulfills the 'cutoff' criteria. Barplot is also available to visualise the data on the 'Publication Plots' tab, which includes all publications per year, stratified by order of authorship (Figure~\ref{fig:read1}). A separate barplot summarises the number of citations per year, and the number of citations per paper per year for an individual author.

\subsection{Network plot}
A unique feature of the 'Super Researcher' app is the network plots which provide insight into the publication culture in the network of researchers who are co-authors with the index author. Each author in the network is represented by a 'node' which is connected to the index author by an 'edge'. The weight of the node and edge is influenced by the number of shared publications between the two authors, which is visualised as the size of the node and the thickness of the edge. Users have the option to find out the network for any selected year or for all years. The name and Scopus ID of the co-authors are showed so that the user can find out more information about them if indicated. Three types of network plots, 'first author network plot', 'last author network plot' and 'co-authors network plot', are available. 

\begin{figure}[!h]
\includegraphics[height=2.18in, width=3.5in]{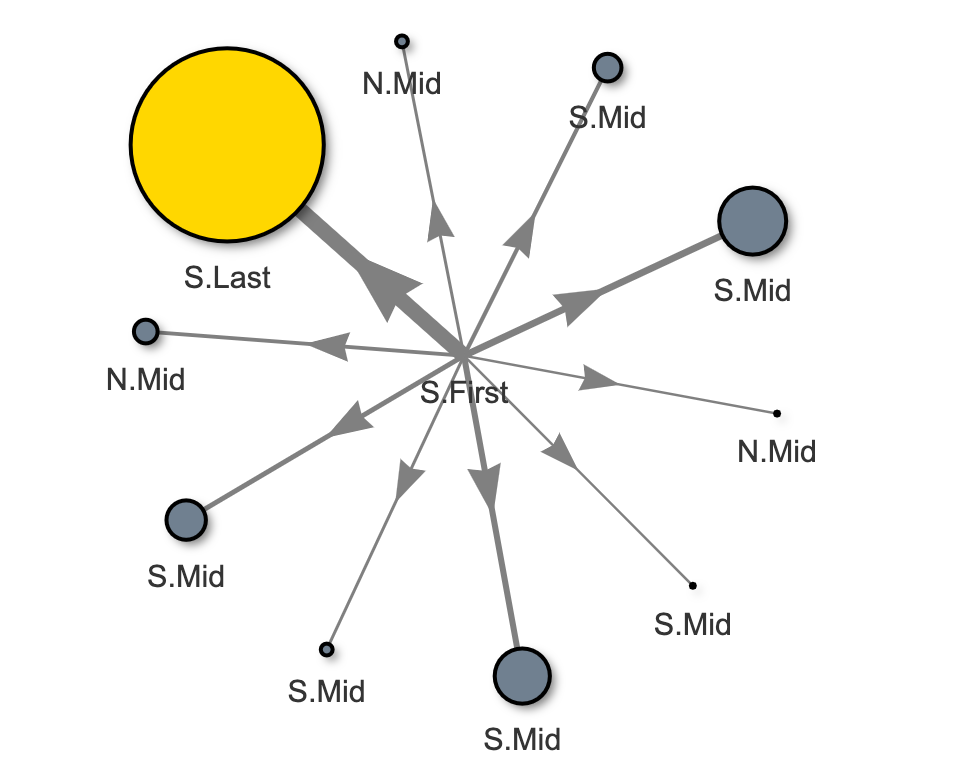}
\caption{First Author Network plot of co-authors with index author as the first author. Prefix 'S' and 'N' represents 'Super Researcher' or 'Normal Researcher' respectively. 'Last', 'Mid' or 'First' indicates the order of authorship. Size and thickness of the node and edge is influenced by the number of shared publications between the two authors. Scopus ID and name of author is available on hovering over the nodes on the interactive app}
\label{fig:network1}
\end{figure}
The first author network plot shows the network of co-authors that the index author worked closely with when he or she is the first author. Co-authors who appear in this network are likely to be from the same research group as the index author, or close collaborators from other groups. Figure~\ref{fig:network1} showcases a network plot of a 'Super Researcher' with the prefix 'S' who is highly connected to a single last author (S.Last). 

\begin{figure}[!h]
\includegraphics[height=2.3in, width=3.5in]{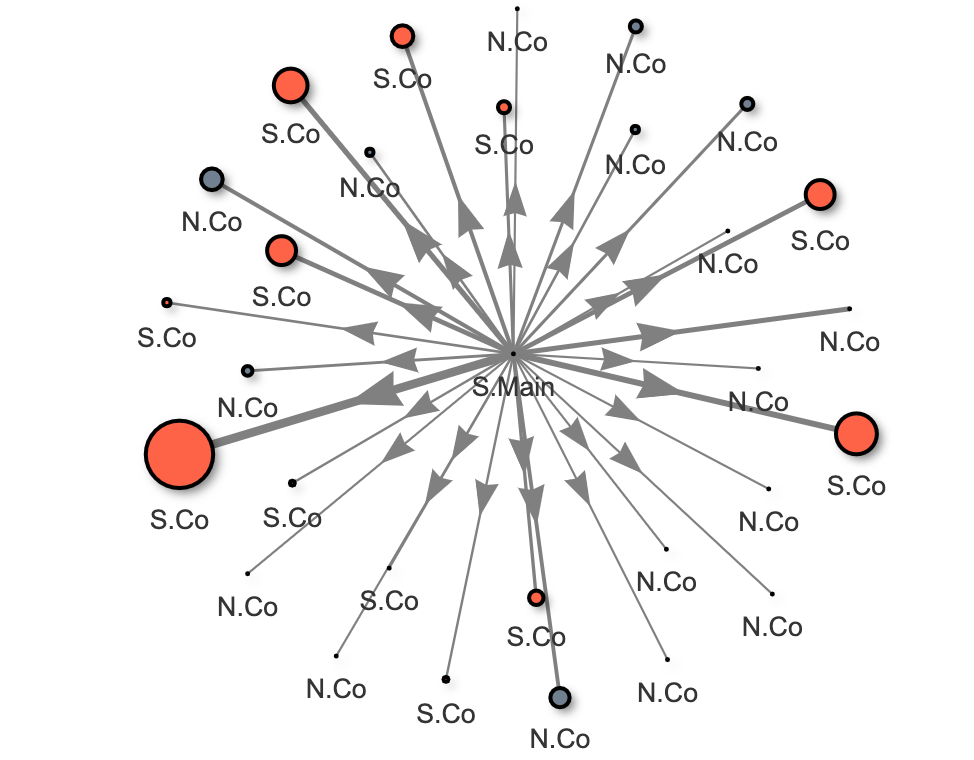}
\caption{Last Author Network plot of co-authors with index author as the last author. Prefix 'S' and 'N' represents 'Super Researcher' or 'Normal Researcher' respectively. 'Last', 'Mid' or 'First' indicates the order of authorship. Size and thickness of the node and edge is influenced by the number of shared publications between the two authors. Scopus ID and name of author is available on hovering over the nodes on the interactive app}
\label{fig:network2}
\end{figure}

The last author network plot shows the network of co-authors when the index author is the last author (Figure~\ref{fig:network2}). This network is more relevant when examining the co-author network of a PI, who will be the last author in most publications. The app algorithm is designed to give priority to coauthors for whom the index author was the last author.

The co-authors network plot shows the network of co-authors that the index author worked with, regardless of the order of authorship. The co-authors in this network include those who work closely with the index author as well as those who are part of a larger research consortium, which the author may contribute to, but is not the main contributor to the research. 

The aim of the network plot is to showcase whether the index author is within a research group with the excessive publication, with a few top collaborators or is publishing excessively on their own. This information can be used as a surrogate of a research group's publication culture, which is more reliable than individual publication data. 

\subsection{Journal information}
\begin{figure}[!h]
\includegraphics[height=2.3in, width=3.5in]{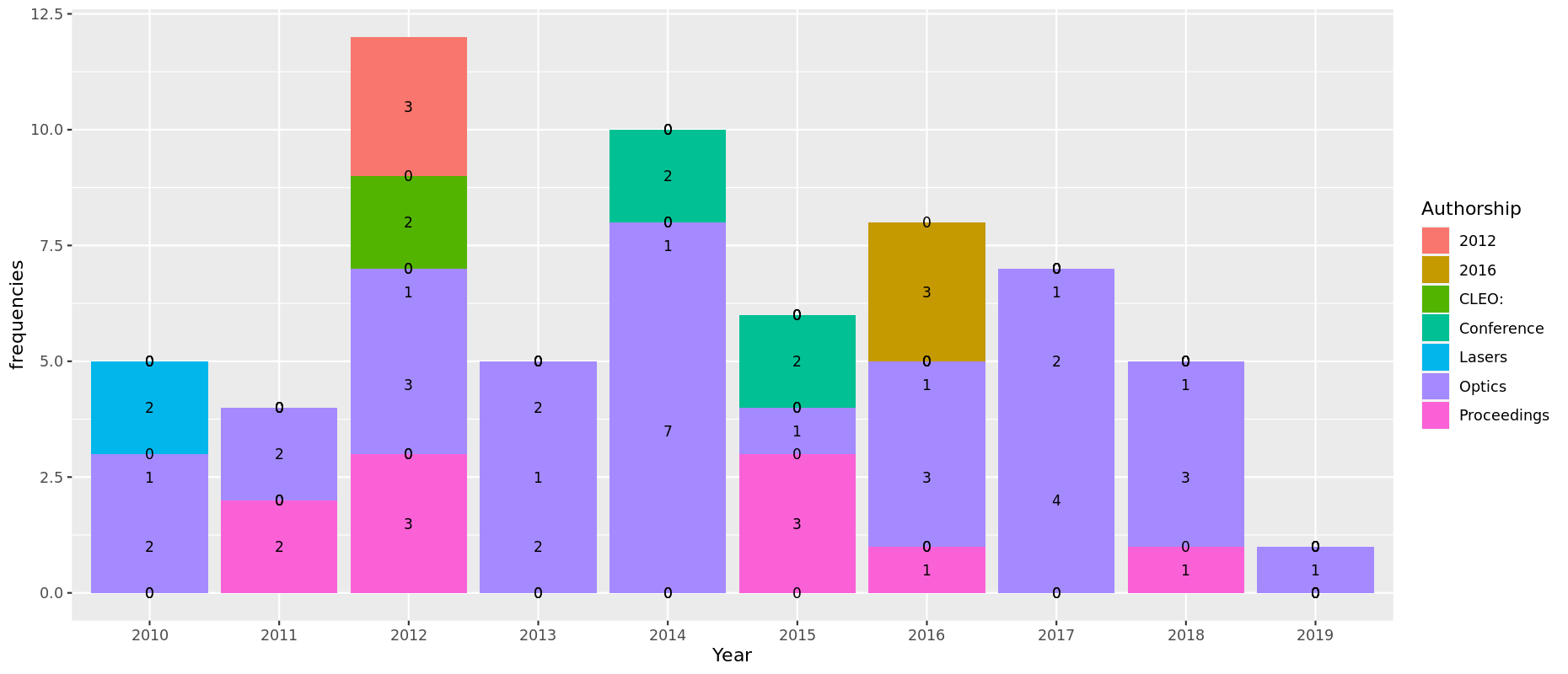}
\caption{Breakdown of top 10 journals which index author published in throughout the years}
\label{fig:journal1}
\end{figure}

The app also provides a breakdown of the top ten journals that the index author published in throughout the years. Besides providing an overview of the quality of publications, this information also showcases whether the author has a habit of publishing his or her work in the same journal(s) repetitively(Figure ~\ref{fig:journal1}). Although there is no hard and fast rule about submitting to the same journals repeatedly, some researchers would prefer to work in a group whose research has a wider acceptance within the scientific community as compared to the same editorial committee of a few journals. 

\section{Methods}

The 'Super Researcher' app uses data available from Scopus to showcase various publication trends about an institution or an individual researcher. At present, due to lack of financial funding, it is not possible to store and analyze data for approximately 17 million researchers from all over the world. Therefore, the app is built in two different instances (local, server). The local instance runs on existing data stored within SQL/NoSQL databases. The server instance directly fetches data from Scopus for each query, which is used for the 'add author' feature. Both instances face different challenges during data collection and analysis. We used Apache Spark with R to handle parallel and complex computations to overcome these challenges.

The local instance uses $get\_author\_info$ function (Algorithm 1) to download author information from Scopus and store them in a HBase database. The algorithm (Algorithm \ref{extraction}) initialises with a parallel computing function \emph{dapply} (Apache Spark function) on each Scopus ID and downloads author information using \href{https://johnmuschelli.com/rscopus/index.html}{rscopus} package (line 2-4). For each year, the algorithm calculates the author-publication numbers as first, mid, and last author (line 5-8). For each publication in that year, it retrieves the names of first, mid, last, and co-authors (line 9-15). It also adds metadata including frequency of publications with all authors (line 16-20). Data from each year are stored separately as a data frame. Finally, it makes a list of the most frequent 30 coauthors, adds publication statistics for them in a data frame, and save data frame to SQL database (line 23-26).

\begin{algorithm}
\caption{Download and Store Author Information}
\label{extraction}
\begin{algorithmic}[1]
\Input List of Scopus ID's $S_{ID}$
\Output Author Data Dataframe $A_{DF}$
\Procedure{Get\textendash Author\textendash Info}{}
\For{each scopus id $sid$ \Pisymbol{psy}{206} $S_{ID}$ }
\State \textbf{dapply}($sid$)
\State data = rscopus::auhtor\_data
\For{each year $y$ \Pisymbol{psy}{206} $data\$year$ }
\State $\alpha$ = first\_author\_pub\_count(data, $sid$, $y$)
\State $\beta$ = mid\_author\_pub\_count(data, $sid$, $y$)
\State $\gamma$ = last\_author\_pub\_count(data, $sid$, $y$)
\State $pubs$ = author\_publications(data, $sid$, $y$)
\For{each publication $p$ \Pisymbol{psy}{206} $pubs$ }
\State $first$ = first\_author\_names($sid$, $y$, $p$)
\State $mid$ = mid\_author\_names($sid$, $y$, $p$)
\State $last$ = last\_author\_names($sid$, $y$, $p$)
\State $co$ = co\_author\_names($sid$, $y$, $p$)
\State Yield($first$, $mid$, $last$, $co$)
\EndFor
\State $first_{meta}$ = frequency(unique($first$))
\State $mid_{meta}$ = frequency(unique($mid$))
\State $last_{meta}$ = frequency(unique($last$))
\State $co_{meta}$ = frequency(unique($co$))
\State Yield($\alpha$,$\beta$,$\gamma$,$first_{meta}$,$mid_{meta}$,$last_{meta}$,$co_{meta}$)
\EndFor
\State $A_{DF}$[$sid\_y$] =c($\alpha$,$\beta$,$\gamma$,$first_{meta}$,$mid_{meta}$,$last_{meta}$,$co_{meta}$)
\State Yield($A_{DF}$)
\EndFor
\State $co_{top}$ = sorted($co_{meta}$)[1:30]
\For{each co\-author id $cid$ \Pisymbol{psy}{206} $co_{top}$}
\State $A_{DF}$[$cid$] = get\_author\_info($cid$)
\EndFor
\State SQLite::save($A_{DF}$)
\EndProcedure
\end{algorithmic}
\end{algorithm}

Once the data about a researcher and his/her most frequent co-authors are available, an extra layer is added to store the relevant information in SQL database from HBase database. Then, three algorithms are developed to analyze and plot researcher publication history, network plot, and frequent journals list.  The publication history algorithm is simple and straightforward. It extracts the data about publication count as first, mid and last author from SQL database and plots them using ggplot2 package ~\cite{ggplot2} (Algorithm \ref{network}). The network plot algorithm is computationally expensive and complex. It initializes with a \emph{dapply} function on selected Scopus ID to extract data from SQL database  (line 1-4). Then, it assigns a label as 'Normal' or 'Super Researcher' based on the number of maximum publications for the selected researchers in any year  (line 5). A new entry is added to the node data frame for the selected researcher as Scopus ID, label, and weight 1 (line 6). It repeats the same process for most frequent coauthors while calculating weight based on their occurrences with the researcher  (line 7-12). Lastly, it uses \href{http://datastorm-open.github.io/visNetwork/}{visNetworkPlot} to show an interactive plot for the researcher (line 13). This algorithm can be modified to extract coauthors data when the index researcher was the first author or author of any order.

\begin{algorithm}
\caption{Researcher Network Plot}
\label{network}
\begin{algorithmic}[1]
\Input Scopus ID's $S_{ID}$, Selected year or year's $Y$
\Output Data frame  with Network plot nodes $N_{node}$
\Procedure{Network\textendash Plot}{}
\For{each scopus id $sid$ \Pisymbol{psy}{206} $S_{ID}$ }
\State \textbf{dapply}($sid$)
\State data = query\_sql($sid$, $y$)
\State label = SR\_test(first\_author\_count$>=$limit)
\State $N_{node}$[$sid$] = c($sid$, label, 1)
\State $coid$ = data\$coauthor\_ids($sid$, $y$)
\For{each coauthor id $cid$ \Pisymbol{psy}{206} $coid$ }
\State datac = query\_sql($cid$)
\State label = SR\_test(first\_author\_count$>=$limit)
\State weight = count($cid$ \%in\% data\$coids[$sid$])
\State $N_{node}$[$cid$] = c($cid$, label, weight)
\EndFor
\State $N_{node}$[$sid$] $<-$ rbind($N_{node}$[$sid$], $N_{node}$[$cid$])
\EndFor
\State return($N_{node}$[$sid$])
\EndProcedure
\end{algorithmic}
\end{algorithm}

The journal plot algorithm makes a list of journals that the index researcher publishes in (Algorithm \ref{journal}). It initializes by applying \emph{dapply} function on each Scopus ID to extract the data from SQL database (line 1-4). Then, it finds the most frequent top 10 journals used by the researcher (line 5). The algorithm iterates for each year and each publication to find the frequency of publication in each journal in that year (line 6-11). Lastly, it uses these frequencies to plot the Journal data using ggplot2 ~\cite{ggplot2} package (line 12).

\begin{algorithm}
\caption{Researcher Journal Plot}
\label{journal}
\begin{algorithmic}[1]
\Input Scopus ID's $S_{ID}$, Length of Journal Name $L$
\Output Journal Plot $J_{Plot}$
\Procedure{Journal\textendash Plot}{}
\For{each scopus id $sid$ \Pisymbol{psy}{206} $S_{ID}$ }
\State \textbf{dapply}($sid$)
\State data = query\_sql($sid$)
\State $pubs$ = table(data\$publications)[1:10]
\State $Y$ = unique(data\$year)
\For{each year $y$ \Pisymbol{psy}{206} $Y$ }
\State $pubs$\$freq[1:10] = 0
\For{each publication $p$ \Pisymbol{psy}{206} 1:length($pubs$) }
\State $pubs$\$freq[$p$] = count(publications=$pubs$[$p$]
\EndFor
\State Yield($pubs$)
\EndFor
\State return(ggplot($pubs$))
\EndFor
\EndProcedure
\end{algorithmic}
\end{algorithm}

The server instance uses similar algorithms to analyze the data and generate various plots. But it queries Scopus every time to extract information instead of the local SQL database. At present, it may take up to a few minutes to download data about a researcher from Scopus. 

\section{Case Study}
In this section, we illustrate how M, a hypothetical postgraduate student who has just completed his MSc degree can utilise the 'Super Researcher' app to gain insight into the publication culture of several research groups before selecting a Ph.D. programme. M views publication as an integral part of a Ph.D. studentship and would like to be within a research group that encourages the publication of high-quality original research work. However, he is also aware of anecdotal experiences from other Ph.D. students who were under immense pressure to publish a high volume of articles every year and would like to avoid that.

M has identified two potential groups which are offering Ph.D. student programme in the area that he is interested in. He would like to learn more about the publication culture of both groups using the 'Super Researcher' app. M sets a minimum of four first author publications per year as the cutoff for 'Super Researcher'.

\begin{figure}[!h]
\includegraphics[height=2.3in, width=3.5in]{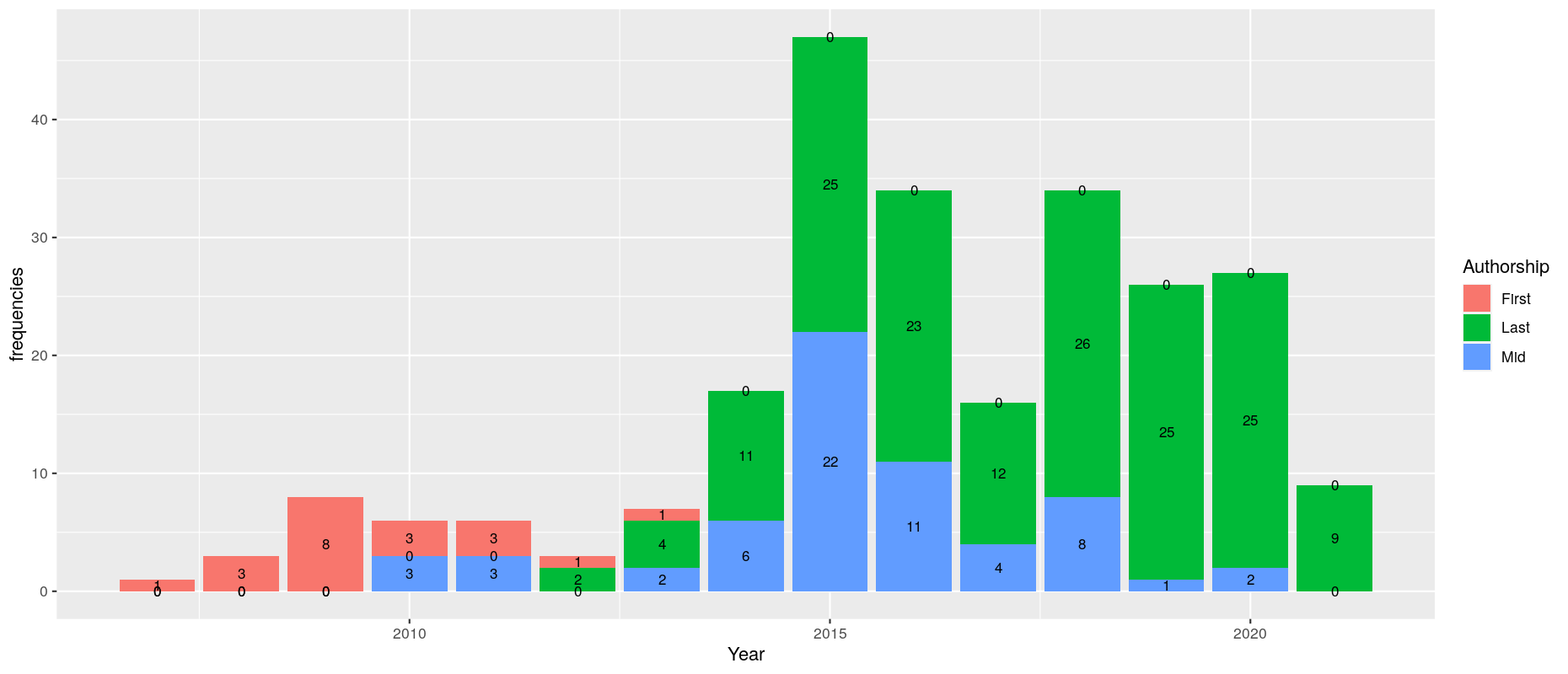}
\caption{Publication plot showcasing the publication number of Prof. A stratified by order of authorship.}
\label{fig:casepub1}
\end{figure}

\begin{figure}[!h]
\includegraphics[height=2.3in, width=3.5in]{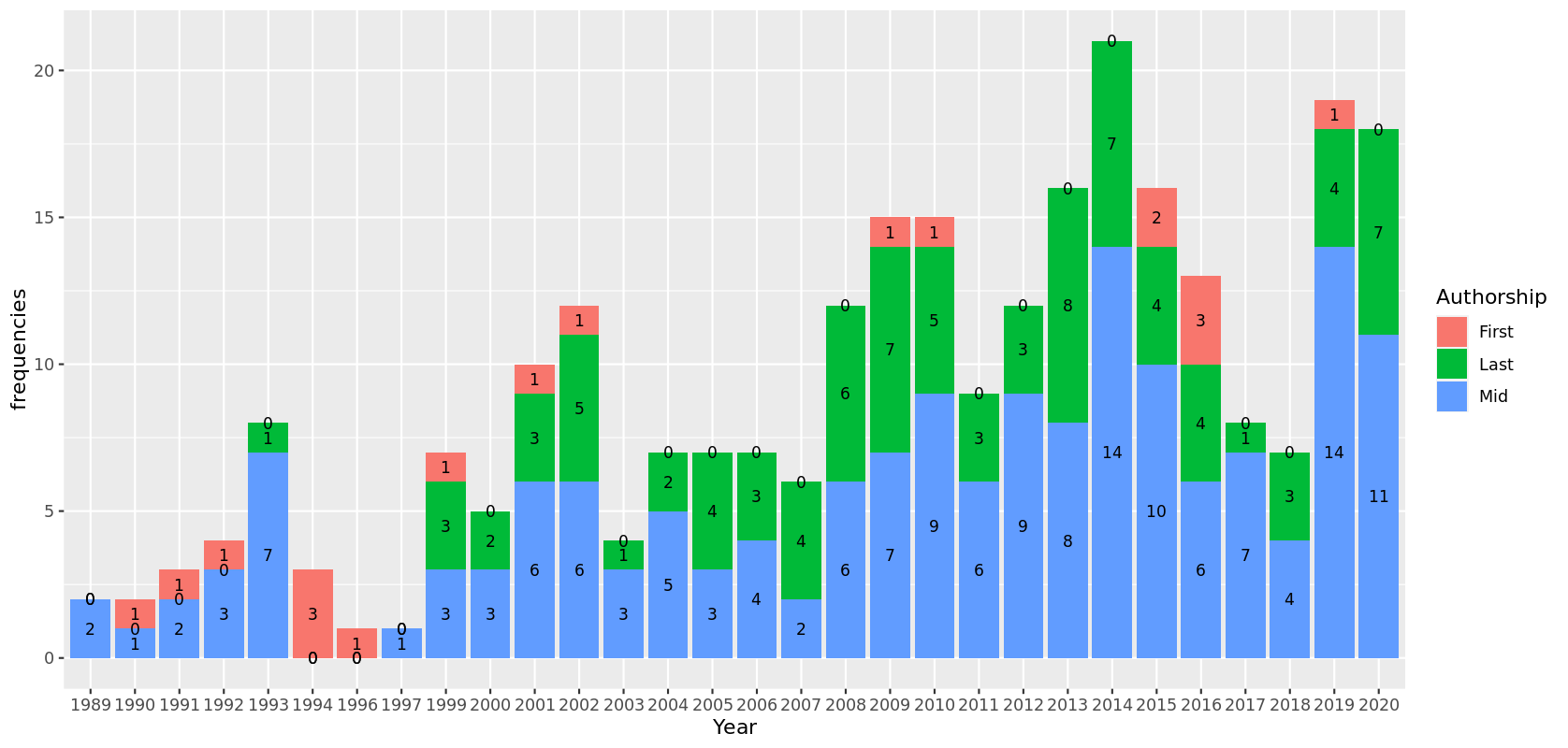}
\caption{Publication plot showcasing the publication number of Prof. B stratified by order of authorship.}
\label{fig:topub1}
\end{figure}

M first looked at the individual author's plot for both professors - Prof. A and Prof B. M discovers that Prof. A was a ’Super Researcher’ in 2009 when he published 8 first-author papers whereas Prof. B was not a 'Super Researcher' based on this cutoff criterion (figure ~\ref{fig:casepub1} and ~\ref{fig:topub1}). In recent years, Prof.A continues to be productive and was averaging 31 publications per year since 2015, the majority as the last author. Prof. B on the other hand averages 13 publications per year in the last six years. The average number of citations per paper per year ranges between 2-41 for Prof. A and 30-550 for Prof. B. 

\begin{figure}[!h]
\includegraphics[height=2.3in, width=3.5in]{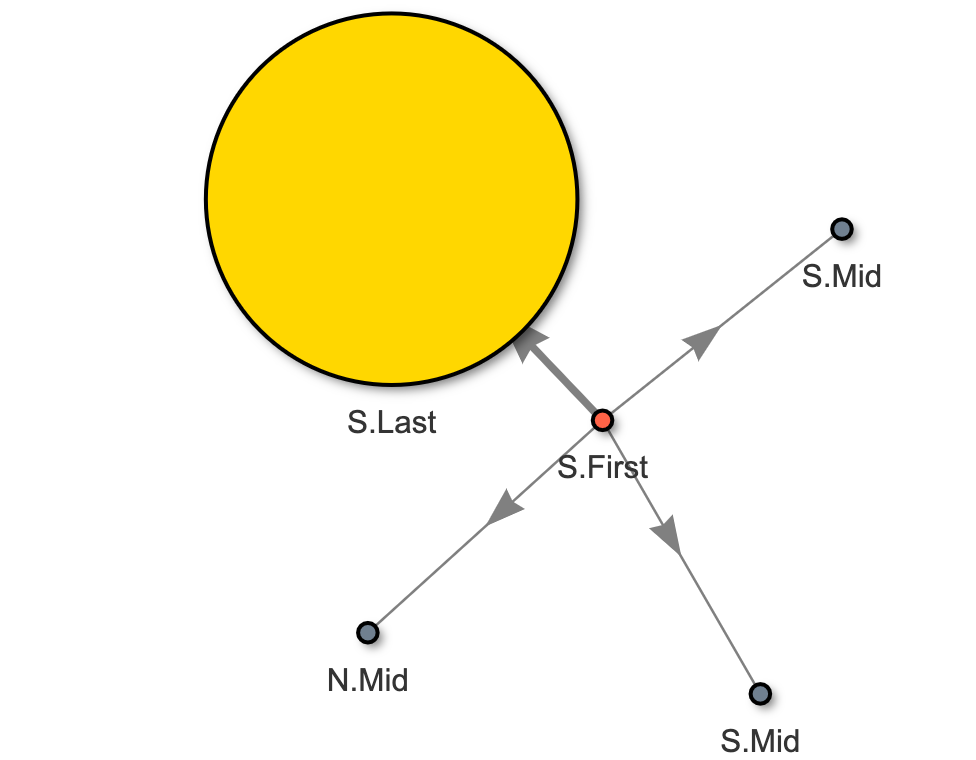}
\caption{Network plot showcasing the network of researchers with co-authorships with Prof. A as first author.}
\label{fig:casenetwork1}
\end{figure}

\begin{figure}[!h]
\includegraphics[height=2.3in, width=3.5in]{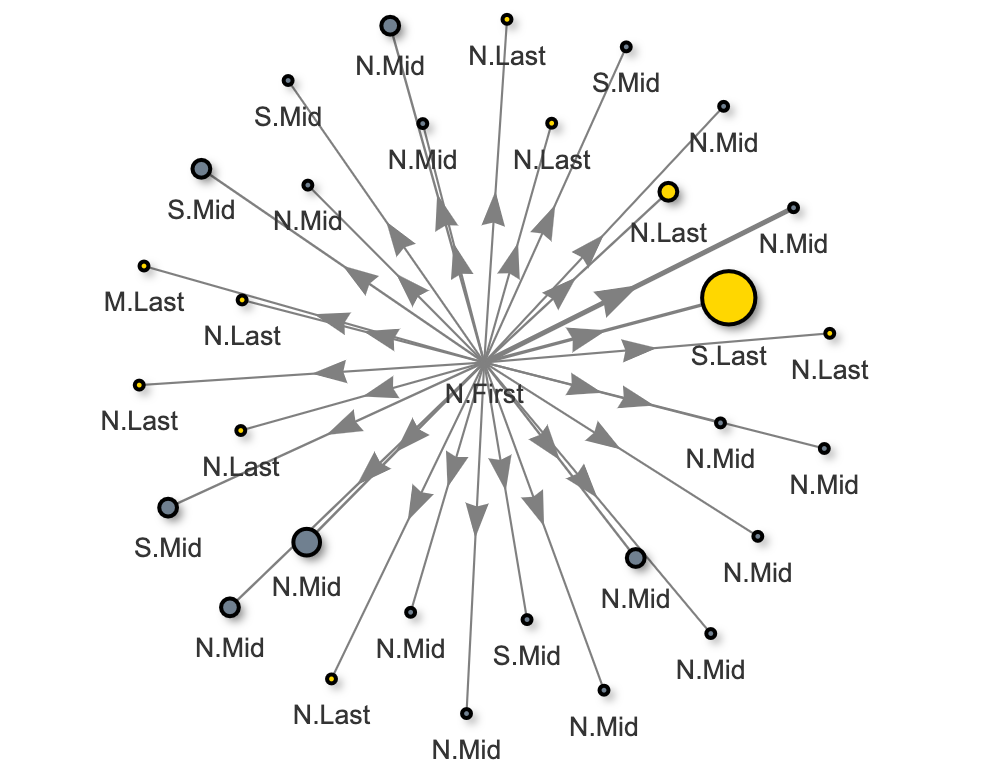}
\caption{Network plot showcasing the network of researchers with co-authorships with Prof. B as first author.}
\label{fig:tonetwork1}
\end{figure}

Next, M examined the network plots further to explore the publication and research network for both groups. When Prof. A is the first author, the network plot shows that he/she published most of his/her articles with the same last author, who is also a 'Super Researcher' (figure~\ref{fig:casenetwork1}). In contrast, when Prof. B is a first author, there is a larger network of co-authors, with several last authors (figure~\ref{fig:tonetwork1}). This is reflective of having worked with different research groups and collaborators, a longer academic career, and having publications since 1989. Within both networks, there are also several other 'Super Researchers'.

\begin{figure}[!h]
\includegraphics[height=2.3in, width=3.5in]{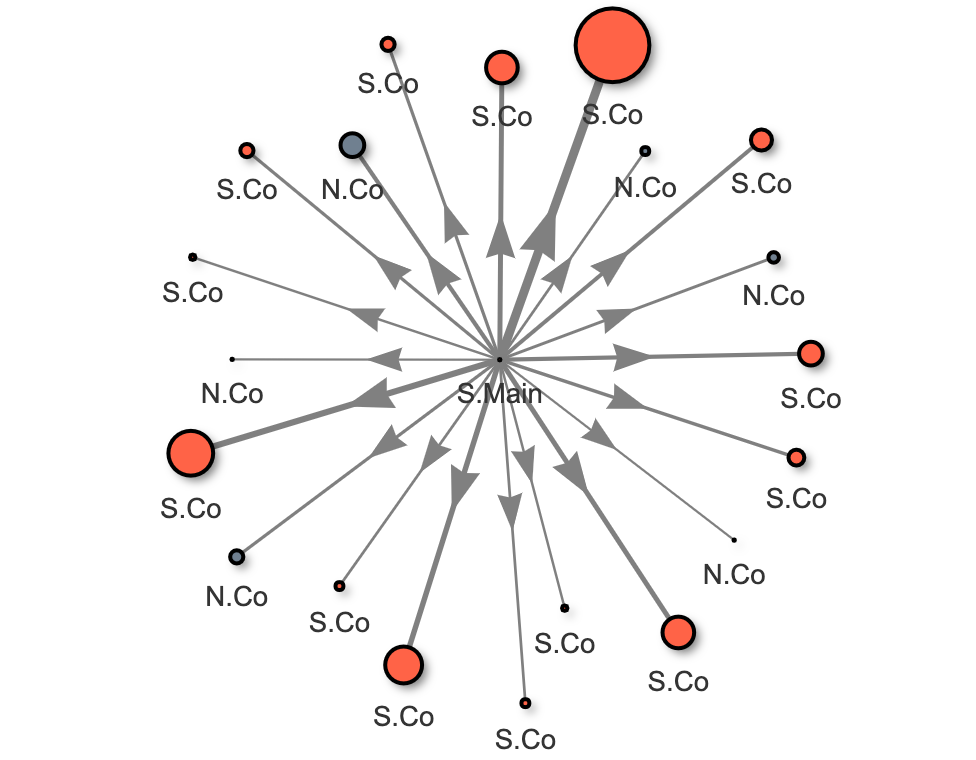}
\caption{Network plot showcasing the network of researchers with co-authorships with Prof.A as last author.}
\label{fig:casenetwork2}
\end{figure}

\begin{figure}[!h]
\includegraphics[height=2.3in, width=3.5in]{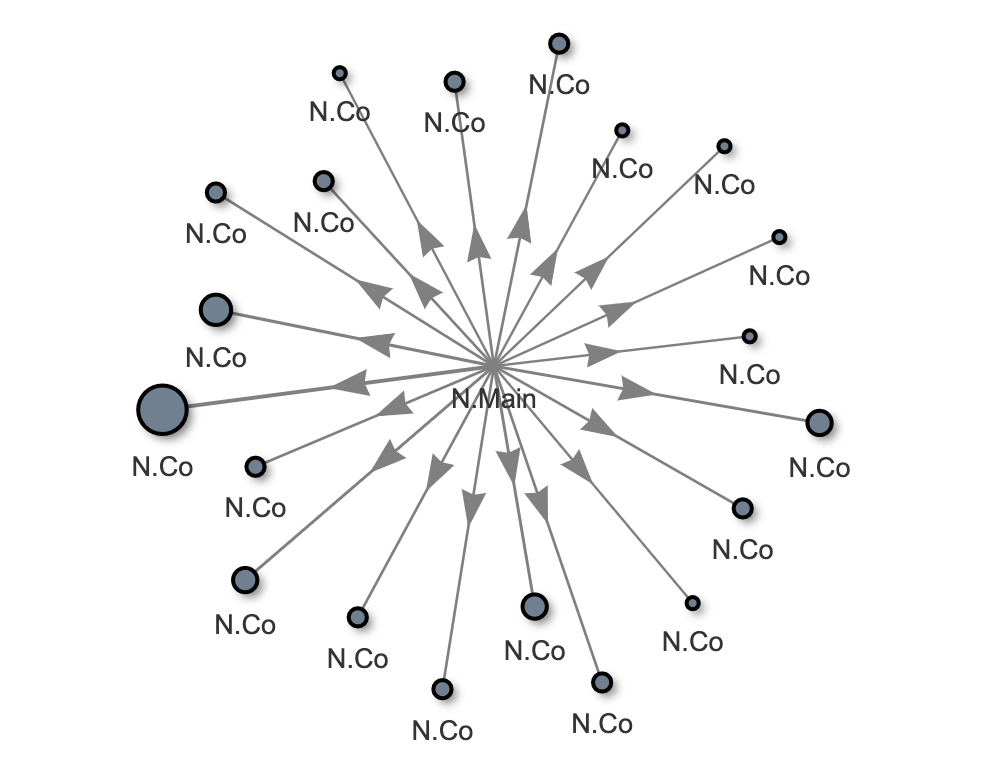}
\caption{Network plot showcasing the network of researchers with co-authorships with Prof.B as last author.}
\label{fig:tonetwork2}
\end{figure}

Next, M examined the network plots for both when they are the last author. M decides to just explore the network plot for the year 2020, which gives an indicator of how many publications M can expect to work on per year in the group. Within Prof. A's network, the majority of his/her co-authors are also 'Super Researchers', publishing more than four first author publications per year (figure~\ref{fig:casenetwork2}). In contrast, Prof. B has no 'Super Researcher' in his/her group (figure~\ref{fig:tonetwork2}). Lastly, M wanted to find out if Prof.A's group publish their work in a high-impact journal. Using the journal plot, M discovered that although publishing at a high volume with a range of citations per paper per year, Prof.A does not have any publication in a high impact journal.

Based on these observations, M can now make a more informed decision regarding joining a new research group for his PhD programme. Prof. A has published extensively and encourages his current group to continue to publish at a high volume, as evident by the presence of many 'Super Researchers' in the network. Despite this, M is concerned that Prof. A's work has not been published in a high impact journal. Prof. B has a much longer career based on his publication history and continues to be highly productive. His/her extensive collaborative network could be advantageous to M. In addition, M might be better suited in an environment where he does not feel under immense pressure to publish numerous first author publication every year.  

\section{Discussion}

Although the pressure to publish is a well-recognized source of stress in academia, a concise method to access summarised publication information is not currently available. Search engines including Scopus, PubMed and Google Scholar provides a list of publications of an author which can be arranged chronologically and filtered based on defined criteria. Various citation metrics including the number of citations, h-index and i10-index are also provided. This data may also be found on visiting an institute's faculty page. However, this data is individualized and is insufficient in providing a comprehensive view of the publication culture of a research group. The 'Super Researcher' app adds on another dimension of information by providing the number of publications per year stratified by order of authorship. More importantly, the network plot presents the most common co-authors associated with the index author and highlights whether these co-authors are 'Super Researchers' themselves. The likelihood that these highly connected authors are from the same research group is high, thus providing a better overview of the group's publication culture. To the best of our knowledge, the 'Super Researcher' app is the only interactive app that is able to provide this information at present. 

What makes a researcher a 'Super Researcher'? In this app, the number of first-author publication(s) is used as the defining criteria for 'Super Researcher'. One of the highly contested issues in the publication is the order of authorship, with two positions associated with the most prestige: the first and the last authors. The first author should be someone who has contributed most to the published work, including writing up the manuscript. The last author, on the other hand, belongs to the senior author in the group, usually, the PI or head of the department ~\cite{mohta2010authorship}. Several issues surrounding co-authorships including 'political', 'ghost' or 'honorary' co-authorships (reviewed by Kornhaber et al. ~\cite{kornhaber2015ongoing}) contributions made by the middle authors inherently difficult to evaluate. For the purpose of informing publication culture and estimating time spent on publication-related work, using the number of the first authorship is therefore the closest approximation. In a research group, researchers are commonly expected to contribute in parts to other publications where they may be the mid authors. As such, approximating workload solely by using first authorships is likely to be an underestimation.

It must be acknowledged that publication culture is but one of the many considerations when choosing a new research group to advance one's career. The 'Super Researcher' app merely provides detailed information on the publication activity of an author without weighing in on the scientific vigour, mentorship quality, faculty support and other important factors in a successful postgraduate programme or research career. The availability of information provided by the 'Super Researcher' app should complement other aspects that are important to the user. In addition, the optimum number of publications per year is akin to Goldilock's dilemma and users must decide for themselves which number is 'just right'.

There are limitations in the pilot phase of this app. At present, due to the quota set by Scopus, readily available author data on the 'Super Researcher' app is limited. To circumvent this, there is an 'Add Author' function for users to search for any researcher. Although adding a new author in the pilot app will take several minutes of processing time, all added authors will remain in the database indefinitely, easing the next user who searches for the same author. This capability allows the 'Super Researcher' database to become user-focused over time while inviting new users to contribute to building the database. In addition, selectively extracting data from Scopus allows the 'Super Researcher's database to maintain a manageable size, reducing data loading time for each query. 

During the course of this work, we have made several interesting observations. Firstly, we encountered a number of authors who have published excessively (as the first author) consistently for several years. Secondly, when probed further, we noticed that 'Super Researcher' tends to exist in networks instead of in isolation. Although these are observations made in a limited sample size and are speculative, we believe that these trends pose some interesting questions. Do researchers inherit publication habits from his/her PI and pass it on to their groups when they become PI? Are institutes aware of or are encouraging a culture of excessive publishing within their faculties? We note a recent publication highlighting the detrimental effect of 'predatory' journals on publishing misleading, poor-quality studies on the whole scientific community ~\cite{machavcek2021predatory}. We believe that institutions can contribute positively in improving the quality of the publication and that the 'Super Researcher' app is ideally suited to examine this further.

\section{Conclusion}
The 'Super Researcher' app is a first of its kind tool to address the vacuum in information on the publication culture of a research group. Using Big Data framework Apache Spark and text-mining methodology, we have created a reliable and interactive app that provides four key matrices: institution data, individual author data, author network plots and journal information. These matrices were selected and designed to aid new researchers to make an informed decision regarding the publication culture of a research group. Although initially designed for an individual researcher, we believe that there are other areas where the 'Super Researcher' app functions are applicable, especially in further interrogation of excessive publication culture that permeates academia.   

\section{Future works}
We are actively seeking collaborations from individuals or groups with available infrastructure to host the 'Super Researcher' app. Please contact the authors for any enquiries.

\section{Acknowledgement}

We want to acknowledge University of Oxford and University of Southampton for providing us with a healthy research environment. We are also thankful to our friends who shared their experiences from immense pressure to publish which motivated the inception of the 'Super Researcher' app.

\ifCLASSOPTIONcompsoc
  \section*{}

\else
  \section*{}
\fi

\ifCLASSOPTIONcaptionsoff
  \newpage
\fi



\bibliographystyle{IEEEtran}
%\bibliography{IEEEabrv,IEEEexample}

%\begin{thebibliography}{1}


%\end{thebibliography}

\bibliography{super_researcher}
%\begin{IEEEbiography}[{\includegraphics[width=1in,height=1.25in,clip,keepaspectratio]{sanjay}}]{Sanjay Rathee}
%\end{IEEEbiography}


%\vfill


\end{document}